\documentclass[aps,prb,twocolumn]{revtex4-1}
\usepackage{graphicx}

\begin{document}

\title{Chlorine insertion and manipulation on the Si(100)-2$\times$1-Cl surface \\ in the regime of local supersaturation}

\author{T. V. Pavlova$^{1,2}$}
 \email{pavlova@kapella.gpi.ru}
\author{V. M. Shevlyuga$^{1}$}
\author{B. V. Andryushechkin$^{1}$}
\author{K. N. Eltsov$^{1}$}
\affiliation{$^{1}$Prokhorov General Physics Institute of the Russian Academy of Sciences, Moscow, Russia}
\affiliation{$^{2}$National Research University Higher School of Economics, Moscow, Russia}

\begin{abstract}

We insert and manipulate a single chlorine atom in chlorine monolayer on a Si(100)-2$\times$1 surface using a scanning tunneling microscope. Two objects were created --- a Cl atom in a groove between two dimer rows, and bridge-bonded Cl on a silicon dimer. Changing the voltage polarity leads to conversion of the objects into each other. Anisotropic movement of the objects at 77 K is mediated by two different diffusion mechanisms: hopping and crowdion-like motion. Insertion of a Cl atom in a groove between two dimer rows leads to the formation of a dangling bond on a third-layer Si atom. At positive sample voltage bias, the first object is positively charged, while the second object can be neutral or negatively charged depending on silicon sample doping.

\end{abstract}

\maketitle

\section{Introduction}

The structure and reactivity of a Si(100) surface interacting with chlorine have been investigated extensively due to the technological importance of this system in microelectronic fabrication \cite{2001Aldao}. Molecular chlorine dissociatively adsorbs on the Si(100) surface forming a 2$\times$1 overlayer, with one Cl atom per dangling bond (DB) at the coverage of 1\,monolayer. The absence of a single chlorine atom on a saturated silicon surface leads to the formation of a DB. As was shown for a hydrogenated Si(100) surface, a DB has isolated electronic states within the band gap of silicon \cite{2013Schofield, 2014Taucer, 2015Labidi}. The strong charge localization on these states allows to use a DB as a single-atom quantum dot \cite{2009Haider}. We will show that a Cl adatom inserted into a saturated Si(100)-2$\times$1-Cl surface (Cl(i)) can also lead to the formation of a DB  on a Si atom of the third layer.

Chlorine insertion into a saturated Si(100)-2$\times$1-Cl surface  (called a supersaturation) can not be achieved by Cl$_2$ adsorption. The dissociation of Cl$_2$ and chlorine insertion occur only on a DB that can be initially present on the surface or produced by photon-activated electron-stimulated desorption (PAESD) \cite{2006Trenhaile} at a temperature of 700\,K \cite{2007Agrawal, 2009Aldao}. At 700\,K, the supersaturated Si(100)-2$\times$1-Cl surface follows an unusual etching pathway, which leads to a smoother silicon surface without regrowth islands compared with a surface morphology in the case of conventional etching \cite{2007Agrawal, 2009Aldao}. In the proposed etching mechanism in the supersaturation regime, two Cl(i) atoms pair up and a whole silicon dimer is removed in the form of two SiCl$_2$ units, cutting the possibility for a single Si atom to diffuse to a regrown island. The authors in Ref.~\cite{2009Aldao} reported the identification of the so-called  bright features (BF) observed by a scanning tunneling microscope (STM)  and associated with two or more Cl(i). Although STM results obtained in Refs. \cite{2007Agrawal, 2009Aldao} made a noticeable contribution to the recognition of the supersaturated  structures on Si(100), the quality of the STM images of BF does not exclude an alternative interpretation. The STM images of Cl(i) on the Si(100)-2$\times$1-Cl surface obtained with atomic resolution in our experiments clearly contradict those obtained with the proposed supersaturation etching mechanism \cite{2009Aldao}.

We present results on the local insertion of single chlorine atoms into the saturated chlorine layer on Si(100) performed by means of atomic manipulation. We have found that local supersaturation can be achieved by the adsorption of the Cl atom from the Cl-functionalized STM tip or by the readsorption of Cl from the chlorine layer producing Cl(i) atoms and Cl vacancies. Our high-resolution STM
data supported by density functional theory (DFT) calculations allowed to associate created objects with two local minima (LM) predicted in the works by de Wijs et al. \cite{1996deWijs, 1998deWijs}: LM1 (a chlorine atom in between dimer rows) and LM2 (a chlorine atom inside a dimer in the bridge configuration). In addition, we have found that LM1 and LM2 can be charged and LM1 has isolated states in the band gap at a DB. This study presents atomic-scale details and dynamics of the surface objects formed on the Si(100)-2$\times$1-Cl surface in the regime of local supersaturation.

\section{Experimental and calculation details}

All the experiments were performed in an ultra-high vacuum setup with a base pressure of 5$\times$10$^{-11}$\,Torr.  STM measurements and atomic manipulations were carried out using a low-temperature scanning tunneling microscope GPI CRYO (SigmaScan Ltd.) operating in the temperature range of 5--300\,K. We used B-doped Si(100) samples (p-type, 1\,$\Omega$\,cm), P-doped Si(100) samples (n-type, 0.1\,$\Omega$\,cm), and Sb-doped Si(100) samples (n-type 0.02\,$\Omega$\,cm). Clean Si(100)-2$\times$1 surface was prepared with a standard thermal annealing procedure including flash heating up to 1470\,K \cite{2020Pavlova}. We used polycrystalline W and Pt-Rh tips for both scanning and manipulation experiments. All STM images presented here were recorded at 77\,K, and voltage was applied to the sample ($U_s$).

Spin-polarized DFT calculations were performed by using the Vienna \textit{ab initio} simulation package (VASP) \cite{1993Kresse, 1996Kresse} with a Perdew-Burke-Ernzerhof (PBE) exchange-correlation functional \cite{1996Perdew} and van der Waals correction developed by Grimme  \cite{2006Grimme}. The Si(100)-2$\times$1 surface was modeled by an eight layer slab with a 5$\times$6 surface unit cell. Chlorine atoms were placed on the upper surface plane of the slab, while the dangling bonds on the lower plane were saturated by hydrogen atoms. The bottom three layers were fixed at bulk positions, while the other silicon layers and chlorine atoms were allowed to relax. The slabs were separated by vacuum gaps of approximately 15\,{\AA}. Reciprocal cell integrations were performed using the 4$\times$4$\times$1 k-points grid. The electronic density of states (DOS) was evaluated using the 16$\times$16$\times$1 k-points grid. STM images were simulated in the framework of the Tersoff-Hamann approximation \cite{1985Tersoff}. The adsorption energy of an adatom was calculated as the difference between the total energy of the surface with the adatom and the total energies of the Si(100)-2$\times$1-Cl surface and the adatom in the gaseous phase.

\section{Results and discussion}
To avoid adsorption of water from the residual atmosphere,  the  adsorption of  molecular chlorine on Si(100) was carried out at elevated temperatures just after the flash heating was switched off. We estimated the sample temperature during Cl$_{2}$  adsorption to be in the range of 370--420\,K. To create the uniform defect free Si(100)-2$\times$1-Cl  surface,  we introduced molecular chlorine onto the sample face with a partial pressure of 10$^{-8}$\,Torr during 100--200\,s.

The next step of our work was the functionalization of the STM tip aiming to put chlorine atoms on the tip. For this purpose, we used an STM lithography procedure developed by us for the creation of pits in the Si(100)-2$\times$1 surface covered by a chlorine mask \cite{2020Pavlova}. During STM lithography, the tip was  moved to the desired place on the surface, the feedback loop was turned off and the sample was supplied with a voltage of about $U_s = +4$\,V at a current of $I_{t}=1$\,nA for 2\,s \cite{2020Pavlova}. As a result, etching pits were formed on the surface with the removal of mostly silicon and partly chlorine atoms, so both Cl and Si atoms could be deposited on the tip.

\begin{figure}[!t]
\includegraphics[width=\linewidth]{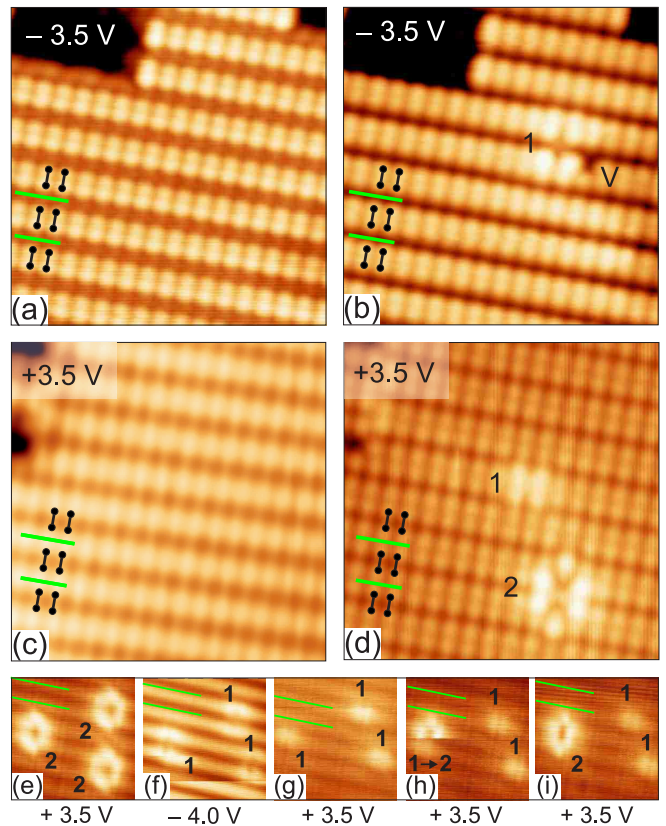}
\caption{\label{fig1} (a)--(d) Insertion of adatoms into the Si(100)-2$\times$1-Cl surface by STM manipulations. Filled state STM images (59$\times$59\,{\AA}$^2$, $U_s =-3.5$\,V, $I_t = 0.5$\,nA) of the Si(100)-2$\times$1-Cl surface (a) before and (b) after a voltage pulse of $-2.0$\,V. Chlorine vacancy (V) and object 1 are formed after the pulse. Empty state STM images (59$\times$59\,{\AA}$^2$, $U_s =+3.5$\,V, $I_t = 0.5$\,nA) of the Si(100)-2$\times$1-Cl surface (c) before and (d) after a voltage pulse of $+2.7$\,V. Objects 1 and 2 are formed on the surface. (e)--(i) Transformations of object 2 into 1 and back in sequentially recorded STM images (48$\times$48\,{\AA}$^2$, $I_t = 1.0$\,nA) while biasing the sample at potentials with both polarities. The slow scan direction proceeded from bottom to top. Silicon dimers are marked with black dumb-bells; grooves between the dimer rows are marked with green lines.}
\end{figure}

To achieve local supersaturation, the STM tip was positioned over the desired location on the surface keeping the scanning parameters unchanged, then the feedback was turned off and a voltage pulse of 1\,ms duration was applied to the sample. The scanning before and after the pulse was carried out at the same voltage polarity as the polarity of the pulse. The pulse of about $-2$\,V usually leads to the creation of one or several similar bright objects and one or several Cl vacancies (Fig.~\ref{fig1}a,b). The center of the bright object in Fig.~\ref{fig1}b is located in the groove between two adjacent dimer pairs. The pulse of about $+3$\,V usually leads to the creation of one or several bright objects. The creation of two different bright objects is shown in Fig.~\ref{fig1}c,d. The center of  object 1 in Fig.~\ref{fig1}d is also located in between four dimers, since the dark stripes in an empty state STM image correspond to lines along the centers of the Si dimers.
The center of  object 2 in Fig.~\ref{fig1}d coincides with the center of the Si dimer.

Scanning objects 2 at high negative voltages ($|U_s| >3$\,V) led to their conversion into objects 1 (Fig.~\ref{fig1}e,f). If the polarity is reversed back to positive, objects 1 are maintained (Fig.~\ref{fig1}g), but they can be converted into objects 2 during the scanning (Fig.~\ref{fig1}h,i).

\begin{figure*}[t]
\includegraphics[width=\linewidth]{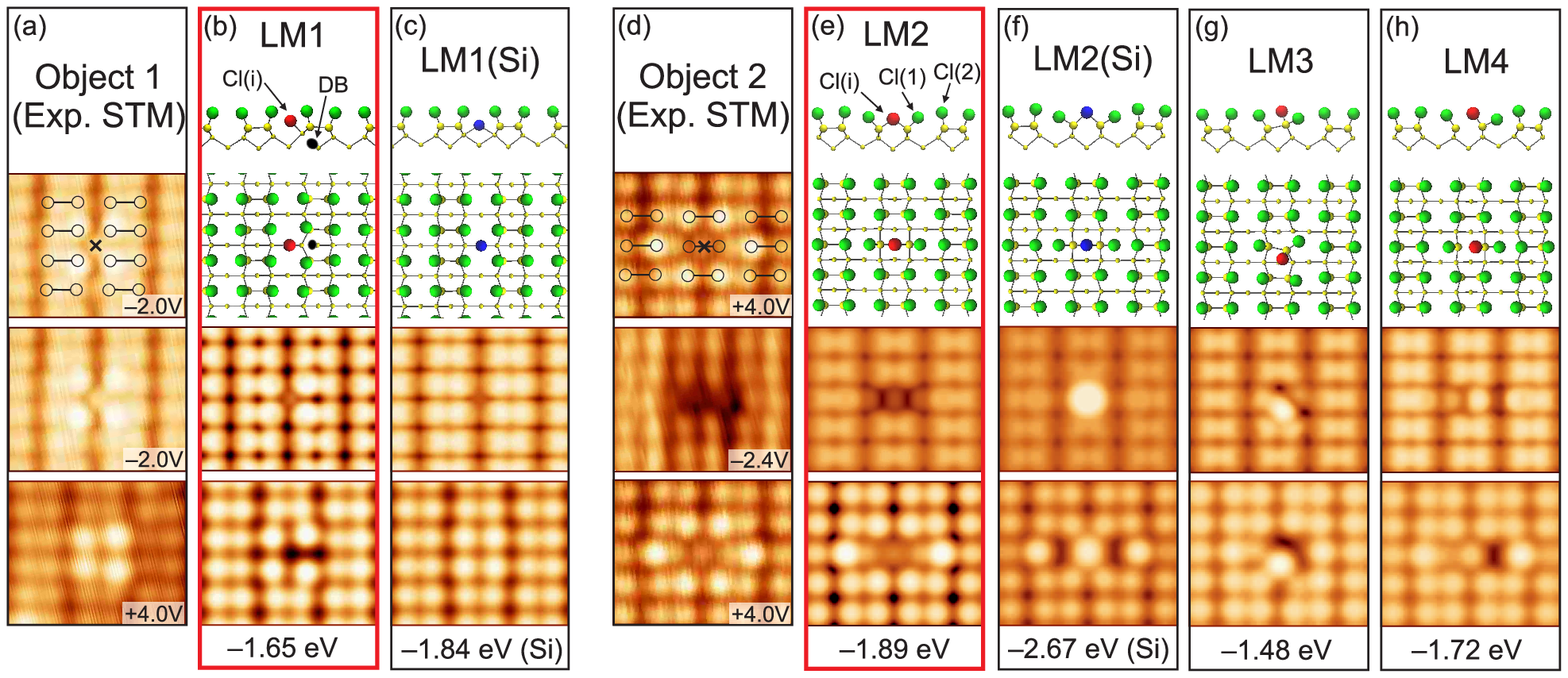}
\caption{\label{fig2} (a) Filled and empty state STM images of object 1. Top panel: the positions of chlorinated Si dimers are marked by dumbbells and the position of the adatom is marked with a cross on the filled state STM image of object 1. (b),(c) Side and top views and simulated filled ($-2$\,V) and empty ($+4$\,V) state STM images of LM1 configuration with Cl (b) and Si (c) adatom. (d) Filled and empty state STM images of object 2. Top panel: the empty state STM image of object 2 with the same notation as in (a). (e)--(h) Side and top views and simulated filled ($-2$\,V) and empty ($+4$\,V) state STM images of LM2 configuration with Cl adatom (e), LM2 configuration with Si adatom (f), LM3 (g) and LM4 (h) configurations with Cl adatom. Chlorine atoms are indicated by green circles, Cl adatom by red circle, Si adatom by blue circle, Si atoms by yellow circles. Adsorption energies of Cl (or Si) adatom are shown at the bottom of the panel for each configuration.}
\end{figure*}

We performed DFT calculations for configurations with Cl and Si adatoms adsorbed in the positions chosen in accordance with positions  of the centers of objects 1 and 2. Figure~\ref{fig2} presents DFT-optimized configurations of Cl and Si atoms on the Si(100)-2$\times$1-Cl  surface and corresponding theoretical STM images shown in comparison with experimental ones.

We indicate the positions of the silicon dimers in  the experimental STM image of object 1  by dumbbells in Fig.~\ref{fig2}a, and we conclude that a Cl or Si adatom is located in the groove between two adjacent dimer pairs.
A chlorine adatom inserted in this position forms a single bond with the Si atom of the second layer (LM1 configuration, Fig.~\ref{fig2}b). A silicon adatom inserted in the groove forms two bonds with Si atoms of the second layer (Fig.~\ref{fig2}c). The main feature that allows us to choose configuration LM1 with a chlorine atom is the asymmetry of the empty state STM image of object 1 (Fig.~\ref{fig2}a). Indeed, the empty state STM image of configuration LM1(Si) is symmetrical due to two identical bonds of the silicon adatom (Fig.~\ref{fig2}c). In contrast, the empty state STM image of configuration LM1 is asymmetric (Fig.~\ref{fig2}b), in good agreement with the experimental STM image of object 1.

In object 2, the adatom is located in the bridge position between two Cl atoms of the Si-dimer (Fig.~\ref{fig2}d). Configuration LM2 contains an additional chlorine atom in the bridge position on the chlorinated Si-dimer (Fig.~\ref{fig2}e). The inserted Cl atom forms two bonds with Si atoms of the same dimer breaking the bond between them. Chlorine atoms initially bonded with Si atoms of the dimer, Cl(1), move apart in the direction across dimer rows, and two adjacent Cl atoms, Cl(2), move higher by 0.14 {\AA} due to repulsive interaction between chlorine atoms. In the filled state STM image, configuration LM2 looks like a dark dimer with three atoms inside (Fig.~\ref{fig2}e), in good agreement with the experimental STM image of object 2 (Fig.~\ref{fig2}d). A silicon adatom in the bridge position also forms two bonds with atoms of the Si dimer, but the simulated STM images (Fig.~\ref{fig2}f) are completely different from the experimental ones (Fig.~\ref{fig2}d). Other local minima LM3 (Fig.~\ref{fig2}g) and LM4 (Fig.~\ref{fig2}h) with the Cl(i) atom forming the SiCl$_2$ unit on one of the Si atoms of the chlorinated dimer give asymmetric simulated STM images that are not similar to those observed in the experiment (Fig.~\ref{fig2}d). Therefore, only STM images of configuration LM2 are consistent with experimental STM images of object 2, and, in addition, configuration LM2 is the most stable among the considered configurations with Cl adatom.

Thus, we can conclude that objects 1 and 2 are formed by Cl insertion into Si(100)-2$\times$1-Cl in LM1 and LM2 configurations from the STM tip, which was therefore functionalized by chlorine before the creation of the objects.

\begin{figure}[h]
\includegraphics[width=\linewidth]{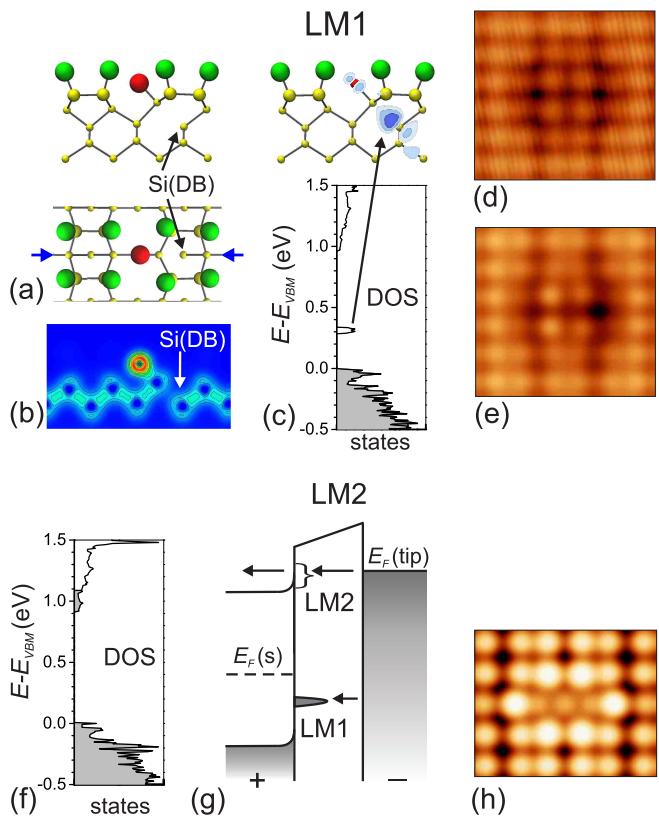}
\caption{\label{fig3} (a) Side and top views of configuration LM1 with Cl(i) adatom. (b) Cross-section of the electron charge density in the plane marked by blue arrows in (a). Red areas indicate larger charge densities, whereas blue areas indicate smaller charge density. (c) Total DOS of LM1 configuration in the vicinity of the band gap. Energy is given relative to the valence band maximum ($E_{VBM}$). Occupied states are filled with gray. Top panel: Integrated Local Density of States (ILDOS) of the unoccupied state in the band gap with isosurface levels of 0.002, 0.004, and 0.008 e/{\AA}$^3$ (from light to dark blue).  (d) Empty state STM image (U$_s =+1.9$\,V, I$_t$ = 1.3\,nA, Sb-doped Si(100) sample) of LM1 with a halo on the Si(100)-2$\times$1-Cl surface due to charge localization on the DB site. (e) Simulated STM image of LM1 at $+1.9$\,V. An electron was added into an enlarged super-cell 10$\times$12 to reproduce the halo. (f) Total DOS of configuration LM2 in the vicinity of the band gap. (g) Band diagram for empty state STM imaging. $E_F$(tip) and $E_F$(s) are Fermi levels of the tip and sample, respectively. $E_F$(s) is indicated by a dotted line since its position depends on the doping level of the sample. At positive sample bias, LM1 can be neutral or negatively charged depending on $E_F$(s) and TIBB, while LM2 is positively charged due to TIBB. (h) Simulated STM image ($+2.0$\,V) of LM2 with one electron removed from a super-cell.}
\end{figure}

Interestingly, the incorporation of a Cl adatom into a chlorine monolayer between dimer rows leads to the formation of a dangling bond on the Si atom of the third layer (Fig.~\ref{fig3}a). The bond between the Si atom of the second layer and the Si atom of the third layers is broken due to Si--Cl(i) bond formation. Figure~\ref{fig3}b shows electron charge-density map of the plane containing a Cl(i) adatom. The absence of charge between the Si atom of the second layer bonded to Cl(i) and the Si atom of the third layer (Si(DB)) corresponds to the breaking of the Si-Si bond.

The electronic density of states (DOS) for configuration LM1 in the vicinity of the band gap is displayed in Fig.~\ref{fig3}c. The presence of isolated electronic states in the silicon band gap is a well-known property of DBs on a hydrogen-terminated Si(100)-2$\times$1 surface \cite{2013Schofield, 2014Taucer, 2015Labidi}. In configuration LM1, there is a level of the DB in the band gap that can hold an electron. The electron probability associated with the unoccupied level is given in the top panel of Fig.~\ref{fig3}c. The DB state is localized, because the highest electron density is found at Si(DB), although the presence of a low electron density on the Cl(i) and neighboring Si atoms indicates a weak coupling between DB and other states. A positive voltage applied to a silicon sample induces upward band bending (tip induced band banding, TIBB) that leads to the increase of the energy of states. If the Fermi level of the sample lies above the level of the DB of configuration LM1 (in the case of n-type samples), the DB state is occupied by an electron. Negative charge on the DB induces the upward band bending effect on the surrounding surface, which is visualized as a "halo" in STM at low positive voltage bias. This effect is well known in the case of the hydrogenated Si(100)-2$\times$1 surface \cite{2009Haider, 2010Bellec, 2013Schofield, 2014Taucer}. Indeed, in the empty state STM image of object 1 at low bias, a halo is clearly seen (Fig.~\ref{fig3}d). The halo is well reproduced in the simulated STM image of the negatively charged LM1 (Fig.~\ref{fig3}e).

On the contrary, LM2 has a level that lies slightly below the conduction band minimum, and, therefore, an electron occupying this level is weakly bonded (Fig.~\ref{fig3}f). At positive sample voltage bias, a weakly bonded electron tunnels from the level at the conduction band minimum into unoccupied energy levels of the sample due to the TIBB (Fig.~\ref{fig3}g). Figure~\ref{fig3}h shows the simulated empty state STM image of LM2, in which one electron was removed from the weakly bonded state. The region near the bridge bonded Cl atom looks brighter (in particular, Cl atoms of the neighboring dimers in the direction along dimer rows), which is in excellent agreement with the experimental empty state STM image of object 2 (Fig.~\ref{fig2}d).

According to Ref.~\cite{1998deWijs}, LM1 is the most stable configuration of Cl(i) followed by LM2 and LM3. However, according to our calculations, LM2 is energetically more favorable than LM1, LM3 and LM4 configurations (Fig.~\ref{fig2}). We explain this discrepancy by using a more accurate calculation method in our case than the calculations of the 90s \cite{1998deWijs}.  Although LM1 is less stable than LM2 in a neutral supercell, LM1 becomes energetically more favorable than LM2  by 0.77\,eV if an electron is added to the supercell. This can be a driving force for switching the adatom from LM2 to LM1 when the voltage bias polarity on the sample changes from positive to negative.

\begin{figure}[h]
\includegraphics[width=\linewidth]{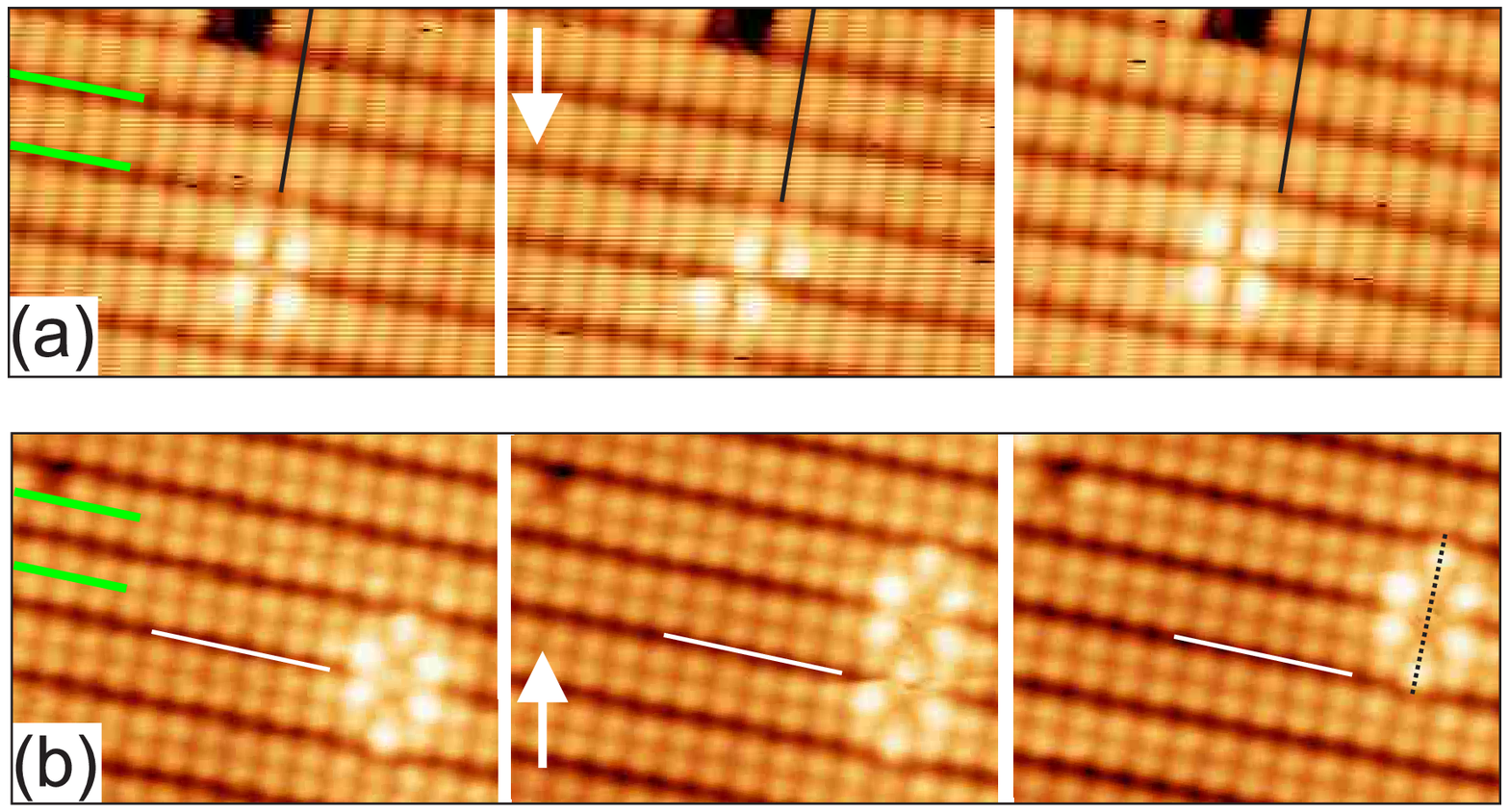}
\caption{\label{fig4} Movement of Cl(i) in the LM1 and LM2 configurations on Si(100)-2$\times$1-Cl. (a) Sequentially recorded filled state STM images (52$\times$40\,{\AA}$^2$, $U_s =-1.7$\,V, I$_t$ = 1.0\,nA) of LM1. (b) Sequentially recorded empty state STM images (52$\times$40\,{\AA}$^2$, $U_s =+3.8$\,V, I$_t$ = 1.0\,nA) of LM2. Five Cl atoms that form a crowdion are marked by black dotted line. Grooves between the dimer rows are marked with green lines. White arrows indicate the direction of scanning. }
\end{figure}

The movement of Cl(i) over the Si(100)-2$\times$1-Cl surface was observed for both LM1 (Fig.~\ref{fig4}a) and LM2 (Fig.~\ref{fig4}b) configurations. In the process of moving, Cl(i) is usually visualized as the first part of bright object, followed by the break through the scan and the second part (or full object) in a new location. The movement of Cl(i) in the LM1 configuration predominantly occurred in the direction of dimer rows (Fig.~\ref{fig4}a). The low energy pathway of Cl(i) diffusion in the LM2 configuration occurs predominantly across dimer rows (Fig.~\ref{fig4}b), although it was predicted to be along dimer rows \cite{1998deWijs}. Therefore, diffusion of Cl(i) in the LM1 and LM2 configurations is sharply anisotropic and occurs in perpendicular directions at 77\,K. The temperature rise from 77\,K to 300\,K leads to the increase of the speed of Cl(i) diffusion and to the change of the diffusion character from anisotropic to chaotic. Note that LM2 and LM1 remain clearly visible in STM at room temperature.

We identified two mechanisms for Cl(i) diffusion in the LM1 and LM2 configurations at 77\,K. The inserted chlorine atom in the LM1 configuration diffuses along the groove between dimer rows (Fig.~\ref{fig4}a) by the hopping mechanism. The inserted chlorine atom in the LM2 configuration and four neighboring Cl atoms marked by a black dotted line in Fig.~\ref{fig4}b are combined into a crowdion --- a linear one-dimensional compression region formed upon the insertion of an additional atom into the lattice \cite{1950Paneth}. A model of a surface crowdion was initially proposed \cite{2003Xiao} to describe ballistic transport of adatoms on strained surfaces. Therefore, the diffusion of Cl(i) in the LM2 configuration in the direction perpendicular to dimer rows (Fig.~\ref{fig4}b) is mediated by such an adatom transport mechanism as the motion of the surface crowdion.

We observed two Cl(i) in LM2 configurations located in the same dimer row and separated by just one dimer, but we did not observe Cl(i) pairing. However, the established etching mechanism in the supersaturation regime involves the diffusion of inserted halogen atoms (Cl or Br) and their pairing in BF due to attraction \cite{2007Agrawal, 2009Aldao, 2016Biswas}. Moreover, bright features were shown in Ref.~\cite{2007Agrawal, 2009Aldao} in filled state STM images, when LM2 looks like a dark dimer or converts to LM1 according to our data. These new details about Cl(i) calls into question the pathway of BF formation as pairing of two inserted halogen atoms in configurations LM2 by hopping along a dimer row, proposed for chlorine \cite{2009Aldao} and bromine \cite{2016Biswas}.

\section{Conclusions}

In summary, we demonstrate that the regime of local supersaturation on the Si(100)-2$\times$1-Cl surface can be achieved by  controllable  atomic  manipulation in STM. Two types of configurations corresponding to the Cl insertion in between dimer rows  (LM1) and in the bridge position (LM2) have been detected and identified in STM images. Configurations LM1 and LM2 can be charged: the charge of LM1 is localized at a DB on a third layer silicon atom and is determined by the Fermi level of the sample and TIBB, and LM2 is positively charged with a positive voltage on the sample. At 77\,K, the mechanism of Cl(i) movement in LM1 and LM2 configurations is completely different and occurs in perpendicular directions. In contrast to the suggestion made in Ref.~\cite{2009Aldao}, we have found a relatively low diffusion rate of inserted chlorine atoms, which allowed them to be visualized in STM over a wide temperature range up to room temperature. We would also like to emphasize the observed modification of the configuration of the inserted chlorine atoms (including their charge state) under the influence of the STM tip, which probably could make it difficult to correctly identify objects LM1 and LM2 in early studies \cite{2007Agrawal, 2009Aldao}.

Our results provide insight into the atomic-scale details and dynamics of the locally supersaturated  Si(100)-2$\times$1-Cl  surface, and they can be utilized in a further study of etching mechanisms of silicon with halogens. In particular, our data can be important in the development of the technology of single silicon atom etching in STM, which is highly desirable for the controllable precise incorporation of phosphorus atoms into a silicon surface for qubits creation \cite{2018Pavlova, 2020Pavlova}.

\begin{acknowledgments}
We   thank  the  Russian  Foundation  for  Basic  Research (RFBR) for funding through the project 20-02-00783. We also thank the Joint Supercomputer Center of RAS for providing the computing power.
\end{acknowledgments}

\bibliographystyle{apsrev4-1}
\bibliography{Pavlova_PRB_rev_arxiv}

\end{document}